\documentclass[a4paper]{revtex4}

\usepackage{amsmath}
\usepackage{epsfig}

\usepackage[latin2]{inputenc}
\usepackage[T1]{fontenc}

\newcommand{\ie}{{\it i.e.}}
\newcommand{\eg}{{\it e.g.}}

\begin{document}

\title{The simplest non-trivial model of chaotic causal dynamics}

\author{Adam J. Makowski}
\email{amak@phys.uni.torun.pl}
\author{Michał  Frąckowiak}
\email{michalf@ncac.torun.pl}
\affiliation{Institute of Physics, Nicholas Copernicus University, 
 ul. Grudziądzka 5/7, 87-100 Toruń, Poland}

\begin{abstract}

The simplest non-trivial model of chaotic Bohmian dynamics is identified. We argue that its most important features can be observed in more complex models, above all, the presumable mechanism of the appearance of chaos in the Bohmian-type dynamical systems.

\end{abstract}

\pacs{05.45.+b, 03.65.-w, 03.65.Sq}

\maketitle

\section{Introduction}

In the de Broglie-Bohm approach to quantum phenomena \citep{b1,b2,b3} particles possess highly non-classical but well-defined trajectories. They are derivable from the guidance equation:
\begin{equation}
\label{m1}
{\mathbf v}=\dot{{\mathbf r}}=\frac{{\mathbf j}}{|\psi|^{2}}=
	\frac{i\hbar}{2m}\frac{\psi\nabla\psi^*-\psi^*\nabla\psi}{|\psi|^{2}}=
	\frac{1}{m}\nabla S,
\end{equation}
where $\psi({\mathbf r}, t) = R({\mathbf r}, t) \exp\left[(i/\hbar) S({\mathbf r}, t)\right]$ is
a generator of the velocity field.

Thus, so-called quantum chaos can be studied via Eq.\ (\ref{m1}) since trajectories naturally exist  in the de Broglie-Bohm mechanics which was proved to be completely equivalent to the standard Copenhagen version of quantum mechanics. The velocity field (\ref{m1}) preserves its definition also within the hydrodynamical formulation of quantum theory.

Solutions of Eq.\ (\ref{m1}) can be very complicated as it was first believed by Bohm himself \citep{b2}. Since the equations can be nonlinear ones, D\"urr et al \citep{b4} concluded that\ldots {\em there is nothing in Bohmian mechanics which would preclude sensitive dependence on initial conditions of}\ldots Bohmian orbits\ldots {\em and hence positive Lyapunov exponents}. The idea of looking for chaos in Bohmian trajectories, and thus in quantum theory, was also mentioned in \citep{b5,b6} and some time later it was actually fulfilled \citep{b7}. Shortly after that a number of authors \citep{b8,b9,b10,b11,b12,b13,b14} found chaotic orbits for various models and entailed wave functions.

A generic feature of Bohmian mechanics is that the phase space volumes are not conserved by the flow, \ie $\nabla \cdot{\mathbf v}$ does not generally vanish. It is expected \citep{b12}, however,  that the volumes are conserved on the average, which means vanishing of $\lim_{T \to \infty} (1 / T)\int_{0}^{T}{\mathbf \nabla} \cdot {\mathbf v}\ {\rm d}\tau$. Very recently we have been successful \citep{b15} in constructing, within the Bohmian mechanics, a model of a Hamiltonian system ($\nabla\cdot{\mathbf v} = 0$), perturbed by disturbance periodic in time, leading to chaotic solutions for some parameters. 

Most studies on quantum chaology mentioned here \citep{b8,b9,b10,b11,b12,b13,b14} were inspired by Parmenter and Valentine's work \citep{b7}, where the system of two non-autonomous equations (or three autonomous), generated from the 2D anisotropic harmonic oscillator wave functions, was integrated. This work opened a number of interesting questions, among them: (i) what are the necessary conditions for the causal chaotic orbits to exist, especially, whether a model-system of equations playing the role of a limit case for other causal models can be identified, (ii) whether the known properties of volume-conserving systems with a time-dependent perturbation can be observed in the Bohmian dynamics as well.

The problems we have just distinguished, though subjectively chosen, are however important for the causal dynamics itself and for the theory of dynamical systems in general. So far, they have not been fully answered. The reason is that the systems based on Eq.\ (\ref{m1}) are much more difficult to deal with than any other conservative or dissipative system studied so far. In this paper we shall therefore try to identify a model being not only the simplest one in the causal dynamics but also revealing the way the chaotic Bohmian orbits are created. In what follows, the answer to the point (i) will be given and we shall also contribute to the point (ii) above. 

\section{Essential equations}

A particularly interesting problem connected with the point (i) above is the following: what is the simplest form of the wave function $\psi({\mathbf r}, t)$ in Eq.\ (\ref{m1}) that still generates chaotic solutions. After having conducted numerical experiments authors of the work \citep{b7} concluded that it is necessary for $\psi({\mathbf r}, t)$ to be a superposition of at least three stationary one-particle states and at least one pair of the states must have mutually incommensurate energy eigenstates. The  wave function of the two-dimensional anisotropic harmonic oscillator used for the study generated a set of two non-autonomous equations with {\em three} control parameters.

Using the same eigenstates we have recently shown \citep{b15} that a linear combination of only {\em two} stationary states is sufficient to obtain chaotic causal trajectories if one of the states is at least double degenerate. The dynamical system derived in such a way had {\em two} control parameters and when they were put equal to each other the system was proved to be completely integrable.

We are going to show that with two properly chosen stationary states dynamical systems can be generated with one or two or four control parameters leading in each case to chaotic behaviour. To this end let us assume ($\hbar = 1,\: m = 1,\: \omega = 1$)
\begin{equation}
\label{m2}
\psi(x,y,t) = \psi_n (x)\psi_n (y)e^{- i E_1 t} + \left[a_0 \psi_k (x) \psi_n (y) + i a_1 \psi_n (x) \psi_k (y)\right]e^{-i E_2 t},
\end{equation}
where $\psi_j$ are solutions of the stationary 1D Schr\"odinger equation and $E_k$ are eigenenergies of the 2D problems. The second stationary state is obviously double degenerate and $a_0$ and $a_1$ are arbitrary real constants. From the definition of the velocity field given in Eq.\ (\ref{m1}) we have
\begin{equation}
\label{m3}
\begin{split}
\dot{x} &= -\frac{a_0 \varphi'(x)\left[\sin(\varepsilon t) + a_1 \varphi(y)\right]}{\left(\cos(\varepsilon t) + a_0 \varphi(x)\right)^2 + \left(\sin(\varepsilon t) + a_1 \varphi(y)\right)^2},
\\
\dot{y} &= \frac{a_1 \varphi'(y)\left[\cos(\varepsilon t) + a_0 \varphi(x)\right]}{\left(\cos(\varepsilon t) + a_0 \varphi(x)\right)^2 + \left(\sin(\varepsilon t) + a_1 \varphi(y)\right)^2},
\end{split}
\end{equation}
where
\begin{equation}
\label{m4}
\varphi(z) = {\psi_k (z) \over \psi_n (z)},\ \ \varepsilon = E_2 - E_1
\end{equation}
and $\varphi'$ means the first derivative with respect to $z$.

From its construction the set (\ref{m3}) has two free parameters $a_0$ and $a_1$. Nevertheless, their number can grow up to four or more, where $E_1$ and $E_2$ are fixed, when some ideas of the so-called supersymmetric quantum mechanics are used. This method is based on using supersymmetry transformations to derive potentials $\hat V$ isospectral to a given one, say $V$, \ie with exactly the same eigenvalue spectrum. For details of the procedure we refer the reader to literature (see \eg \citep{b17}). Here we are only going to present some formulae ready to use.

Let $\{\psi_n (x)\}$ and $E_n$ be respectively eigenstates and eigenenergies of a Hamiltonian with potential $V(x)$. Define a function $I(x) = \int_{-\infty}^x |\psi_0 (z)|^2 {\rm d}z$, where $\psi_0 (z)$ is the ground (nodeless) state. Then, for properly chosen values of the real parameter $\lambda$, we can generate \citep{b17} a new potential $\hat V(x;\lambda) = V(x) - ({\rm d}^2 / {\rm d}x^2)\ln[I(x)+\lambda]$ with $q=0,1,2\ldots$ less bound states than $V(x)$ has. Now the eigenfunctions $\hat\psi(x;\lambda)$ corresponding to $\hat V(x;\lambda)$ in the strictly isospectral case of $q=0$ with $\hat E_n = E_n$ have the form
\begin{equation}
\label{m5}
\hat\psi_{n+1}(x;\lambda)=\psi_{n+1}(x) + {1 \over 2} \left({1 \over E_{n+1} - E_0}\right){I'(x) \over I(x) + \lambda}\left({{\rm d} \over {\rm d}x} - {\psi'_0(x) \over \psi_0(x)}\right)\psi_{n+1}(x),
\end{equation}
where $n=0,1,2\ldots$ and $\lambda > 0$ or $\lambda < -1$ and the prim denotes the  first order derivative. Now, the normalized wave function $\hat\psi_0$ of the ground state reads
\begin{equation}
\label{m6}
\hat\psi_0 (x;\lambda)={\sqrt{\lambda(1+\lambda)}\over I(x)+\lambda} \psi_0 (x).
\end{equation}
In the limit of $\lambda \to \infty$ the ``new'' functions $\hat\psi$ and potentials $\hat V$ reduce to the ``old'' ones.

The procedure sketched above can be further generalized and one can construct \citep{b18} an infinite number of isospectral families $\hat V(x;\lambda_1, \lambda_2,\ldots)$ having identical bound-state energies.

Since the velocity field in Eq.\ (\ref{m1}) is a functional of wave function we can modify properties of the dynamical system as in Eqs.\ (\ref{m3}), based on two stationary states, by increasing the number of control parameters. In the simplest case, instead of Eq.\ (\ref{m2}), we can now propose
\begin{equation}
\label{m7}
\begin{split}
\hat\psi_{\lambda\mu}(x,y,t) = &\hat\psi_n (x;\lambda)\hat\psi_n (y;\mu)e^{- i E_1 t} + \\ &+\left[a_0 \hat\psi_k (x;\lambda) \hat\psi_n (y;\mu) + i a_1 \hat\psi_n (x;\lambda) \hat\psi_k (y;\mu)\right]e^{-i E_2 t},
\end{split}
\end{equation}
where $\mu>0$ or $\mu<-1$. With the replacement $\psi \to \hat\psi_{\lambda\mu}$ we have from Eq.\ (\ref{m1}):
\begin{equation}
\label{m8}
\begin{split}
\dot{x}&=-\frac{a_0 \hat\varphi'_\lambda (x)\left[\sin(\varepsilon t) + a_1 \hat\varphi_\mu (y)\right]}{\left(\cos(\varepsilon t) + a_0 \hat\varphi_\lambda (x)\right)^2 + \left(\sin(\varepsilon t) + a_1 \hat\varphi_\mu (y)\right)^2},
\\
\dot{y} &= \frac{a_1 \hat\varphi'_\mu (y)\left[\cos(\varepsilon t) + a_0 \hat\varphi_\lambda (x)\right]}{\left(\cos(\varepsilon t) + a_0 \hat\varphi_\lambda (x)\right)^2 + \left(\sin(\varepsilon t) + a_1 \hat\varphi_\mu (y)\right)^2},
\end{split}
\end{equation}
where
\begin{equation}
\label{m9}
\hat\varphi_\lambda (x) = {\hat\psi_k (x;\lambda) \over \hat\psi_n (x;\lambda)},\ \ \hat\varphi_\mu (y) = {\hat\psi_k (y;\mu) \over \hat\psi_n (y;\mu)}.
\end{equation}
Of course as $\lambda \to \infty$ and $\mu \to \infty$ the set of Eqs.\ (\ref{m8}) reduces to that given in Eqs.\ (\ref{m3}).

Since for a single stationary state or for a linear combination of two non-degenerate stationary states dynamical systems resulting from Eq.\ (\ref{m1}) can be easily proved to be non-chaotic, we conclude that {\em equations (\ref{m8}) or (\ref{m3}) are the simplest non-trivial systems within the Bohmian mechanics with possibly chaotic dynamics}. At this point it should be mentioned that chaotic behaviour can be generated from Eq.\ (\ref{m1}) even for just one stationary state \citep{b19}, it is necessary, however, to deal with two particles.

\section{The simplest model of chaotic causal dynamics}

We are now ready to identify the model that could play the role of a limiting case of more complex models of Bohmian dynamics and still revealing their non-trivial properties.

To this end let us observe that Eqs.\ (\ref{m3}) and (\ref{m8}) have the same form and the latter were derived to show a possibility of introducing additional control parameters to the system under consideration. Thus, for the time being, we may restrict ourselves to Eqs.\ (\ref{m3}).

In the simplest non-trivial case we can choose
\begin{equation}
\label{m10}
\varphi(z) = C z,\ \ \varepsilon = 1
\end{equation}
in Eqs.\ (\ref{m4}), which follows from using two lowest states of 1D harmonic oscillator, \ie,  $n=0$, $k=1$ and $\psi_0 (z) = A \exp(-(1 / 2) z^2)$, $\psi_1 (z) = B z \exp(-(1 / 2) z^2)$. Due to the relation $C = B / A$, for normalized states $C = \sqrt{2}$. Since the constant can be absorbed by $a_0$ and $a_1$ we take for simplicity $C = 1$. Now $\psi(x,y,t)$ in Eqs.\ (\ref{m2}) is the state representing 2D isotropic oscillator with eigenenergies $E_1 = 1$ and $E_2 = 2$.

In the case of $a_0 = a_1 = a$, since the integral of the motion $C(x,y,t) = M - a^2 \ln{M} - 2 a (x \cos{t} + y \sin{t})$, where $M = (\cos{t} + a x)^2 + (\sin{t} + a y)^2$ exists, Eqs.\ (\ref{m3}) do not generate chaotic solutions. Moreover, we also have ${\partial \dot x / \partial x} + {\partial \dot y / \partial y} = 0$ which means conservation of the phase space ``volume''. It can be also observed that a function $H(x,y,t)$ exists such that $\dot x = - {\partial H / \partial y}$ and $\dot y = {\partial H / \partial x}$ with $H$ defined as $H = (1 / 2) \ln{M}$.

When $a_0 \ne a_1$ Eqs.\ (\ref{m3}) are no more integrable and the very detailed discussion in \citep{b15} showed that the system of equations can be transformed into the form of a Hamiltonian autonomous system with a periodic non-Hamiltonian perturbation. Then, the method of Melnikov function can also be used to prove formally the existence of chaotic orbits.

The model discussed here is, as yet, the only known model of Bohmian dynamics with the above properties. Presumably no similar case with such unique features can be found. To show that it can serve as a reference model for other models we have prepared a sequence of stroboscopic maps in Fig.\ \ref{f1} for a few pairs of $a_0 \ne a_1$ such that $a_0 = 1.015 A$ and $a_1 = 0.985 A$ with $A = \infty, 100, 10, 4, 2, 1$.

The first plot in Fig.\ \ref{f1} represents a circle which is then deformed into a curve formed of two loops crossing a single point. When $a_0 = a_1$ both the homoclinic orbit of that shape and the hyperbolic point can be found formally as shown with full particulars in \citep{b15}. Here we show instead for $A = 10$ a thin stochastic layer appearing in the vicinity of the orbit spreading out in the area surrounding the point. The smaller the value of $A$ is, the faster the homoclinic orbit breaks up and the fully developed chaotic trajectory can be attributed to the picture for $A = 2$. We have estimated the largest Lyapunow exponent for the trajectory with $x(0) = 1.98$, $y(0) = 0$ and $0 < t < 200000$, $\Delta t = 0.001$, as $\lambda_{max} = 0.06$.
 
\section{Comparison with more intricate models}

We shall now deviate from the above model to show that other more complicated models reveal similar behaviour. Two completely different ways of doing such modifications are briefly sketched below.

\subsection{An isospectral modification}

\begin{figure}
\centering
\includegraphics{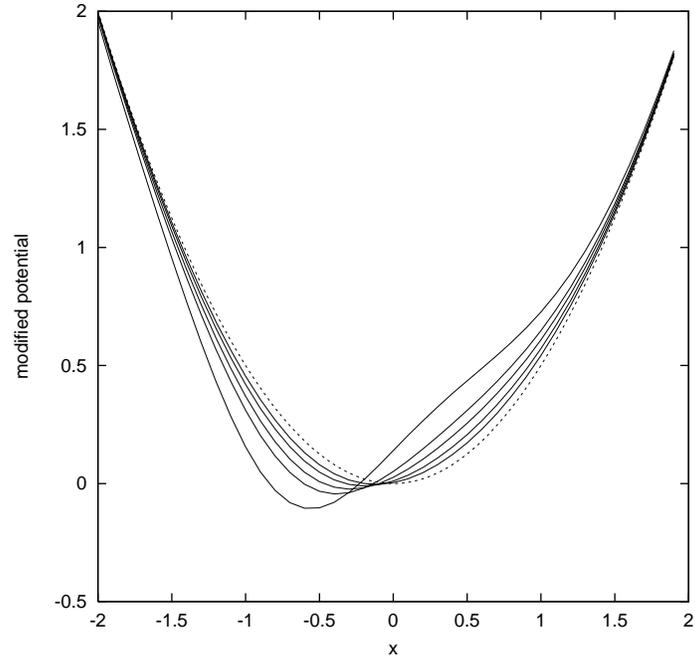}
\caption{
The plot of the modified potential $\hat V_\lambda (x)$ as given in Eq.\ (\ref{m12}) for $\lambda = 1, 2, 3, 5, 9$ (full lines). The smaller the value of $\lambda$ is the larger is the deviation of $\hat V_\lambda (x)$ from the harmonic oscillator potential $V(x) = (1/2) x^2$ or $\hat V_\infty (x)$ (dashed line).
}
\label{f2}
\end{figure}

We can destroy the rotational symmetry of the 2D harmonic oscillator potential of the model of Section 3 and simultaneously preserve the same eigenvalues spectrum of the new potential. To this end the functions $\psi_0 (z)$ and $\psi_1 (z)$, with $z = x$ or $y$, are used in Eqs.\ (\ref{m5}) and (\ref{m6}) and will get $\hat\psi_0 (x;\lambda)$, $\hat\psi_1 (x;\lambda)$ and $\hat\psi_0 (y;\mu)$, $\hat\psi_1 (y;\mu)$. Finally, from Eqs.\ (\ref{m9}), we have:
\begin{equation}
\label{m11}
\begin{split}
\hat\varphi_\lambda (x) = {I(x) + \lambda \over \sqrt{\lambda(1+\lambda)}}\left(x+{e^{-x^2} \over 2\sqrt{\pi}(\lambda + I(x))}\right),
\\
\hat\varphi_\mu (y) = {I(y) + \mu \over \sqrt{\mu(1+\mu)}}\left(y+{e^{-y^2} \over 2\sqrt{\pi}(\mu + I(y))}\right),
\end{split}
\end{equation}
where $I(z)=\pi^{-1/2}\int_{- \infty}^z e^{-u^2}{\rm d}u$. Once more the factor of $\sqrt{2}$ is absorbed by the expansion coefficients in Eq.\ (\ref{m7}). The functions $\hat\psi_0 (x;\lambda)\, \hat\psi_0 (y;\mu)$ and $\hat\psi_0 (x;\lambda)\, \hat\psi_1 (y;\mu)$ correspond to the same previous energies $E_1 = 1$ and $E_2 = 2$, and hence again $\varepsilon = 1$. The new potential has now the form
\begin{equation}
\label{m12}
\begin{split}
\hat V_{\lambda \mu}(x,y) =& \hat V_\lambda (x) + \hat V_\mu (y) =\\
=&  {1 \over 2}(x^2 + y^2) + {2 \pi^{-1/2} x e^{-x^2}\left(\lambda + I(x)\right) + \pi^{-1}e^{-2x^2} \over \left(\lambda + I(x)\right)^2} + \\
&+  {2 \pi^{-1/2} y e^{-y^2}\left(\mu + I(y)\right) + \pi^{-1}e^{-2y^2} \over \left(\mu + I(y)\right)^2} 
\end{split}
\end{equation}
and in the limit of $\lambda \to \infty$ and $\mu \to \infty$ it reduces to the usual 2D oscillator potential. For the sake of comparison, its 1D image is presented in Fig.\ \ref{f2}. The lower the value of $\lambda$ is, the more  $\hat V_\lambda (x)$ deviates from its partner potential $V(x) = (1 / 2) x^2$. Similarly, $\hat\varphi_\lambda(x)$ in Eqs.\ (\ref{m11}) deviates from $\varphi(z)$ in Eq.\ (\ref{m10}). Thus, the model of Section 3 is a limiting case of the one introduced above for asymptotic values of $\lambda$ and $\mu$. Only in this limit with $a_0 = a_1$ the latter is obviously integrable. In other cases, for any combination of the four parameters $a_0$, $a_1$, $\lambda$, $\mu$, a chaotic solution can always be found. The components of the velocity field ${\mathbf v}(x,y,t)$ are now determined by Eqs.\ (\ref{m8}) and (\ref{m11}). Formal calculations show that ${\mathbf \nabla}\cdot {\mathbf v}$ is now changing with time and tends to zero only in the above limit. Nevertheless, the scenario of approaching a chaotic orbit seems to be very similar to that observed for the model given by Eqs.\ (\ref{m10}) and (\ref{m3}). This is clearly visible in Fig.\ \ref{f3}.

The first two plots are about the same as in Fig.\ \ref{f1} and for smaller values of $a$ the stochastic layers appear to be a little more pronounced than those in the corresponding pictures in Fig.\ \ref{f1}. We have expected this as a consequence of the lack of the rotational symmetry of the isospectral Hamiltonian (cf. Fig.\ \ref{f2}).

\subsection{Particle in a restricted space}

\begin{figure}
\centering
\includegraphics[scale=0.75]{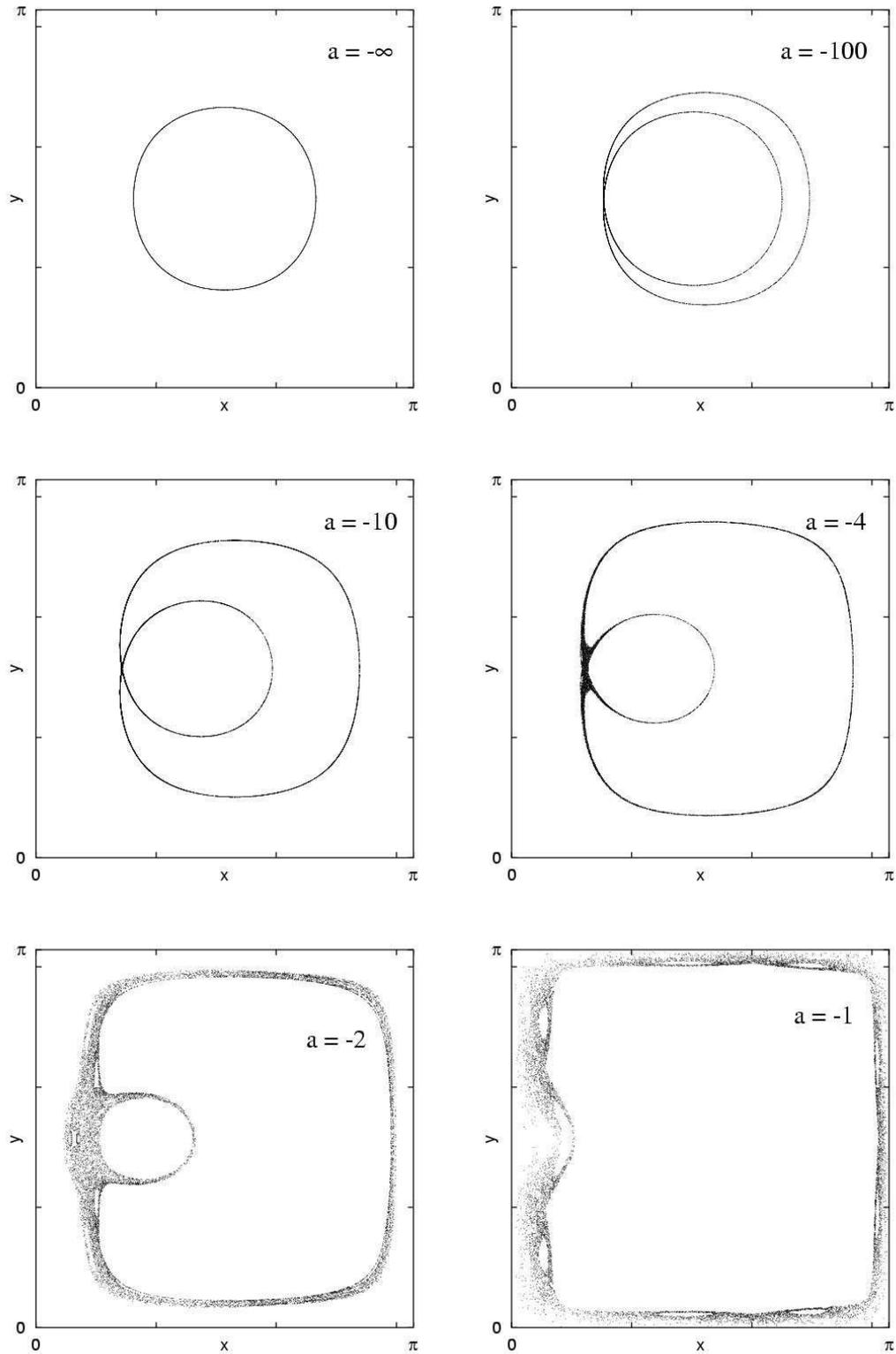}
\caption{
Stroboscopic maps of the period $(4/3) \pi$ for an infinite square well model described by Eqs.\ (\ref{m3}) and (\ref{m13}). The time interval is in each case $0 < t < 50000$ and $\Delta t = 0.001$.
}
\label{f4}
\end{figure}

The second modification of our Eqs. (\ref{m10}) and (\ref{m3}) uses the simplest model with a non-linear eigenvalue spectrum. It is a square well with $x \in [0,\pi]$, $y \in [0,\pi]$ and the normalized wave functions $\psi_n = \sqrt{(2 / \pi)} \sin{(nz)}$ and energies $E_n = (1 / 2) n^2$, $n=1,2\ldots$. Then, for $n=1$ and $k=2$ in Eq.\ (\ref{m2}) we use in Eqs.\ (\ref{m3})
\begin{equation}
\label{m13}
\varphi(z) = 2 \cos z,\ \ \varepsilon = {3 \over 2}.
\end{equation}
In our numerical calculations the factor of 2 is included in the constants $a_0$ and $a_1$ in Eqs.\ (\ref{m3}). Now, Eqs.\ (\ref{m3}) are integrated for a number of values of $a_0 = a_1 = a$ and the results are presented in Fig.\ \ref{f4}. The first member ($|a| >> 1$) of the family of pictures is not a circle but a closed curve perfectly approximated by the relation $\Delta = \sin{x} \cdot \sin{y}$, where $0 \le \Delta \le 1$. For smaller values of the parameter $|a|$ we can again observe the formation of a homoclinic orbit with the shape as in Fig.\ \ref{f4} for $a = -10$. When decreasing $|a|$ the orbit breaks up and the stochastic layer becomes well pronounced.

The dynamics here is very similar to that in our model of Section 3 and this again suggests considering it as the simplest non-trivial model of chaotic Bohmian dynamics.

\section{Conclusion}

\begin{figure}
\centering
\includegraphics[scale=0.75]{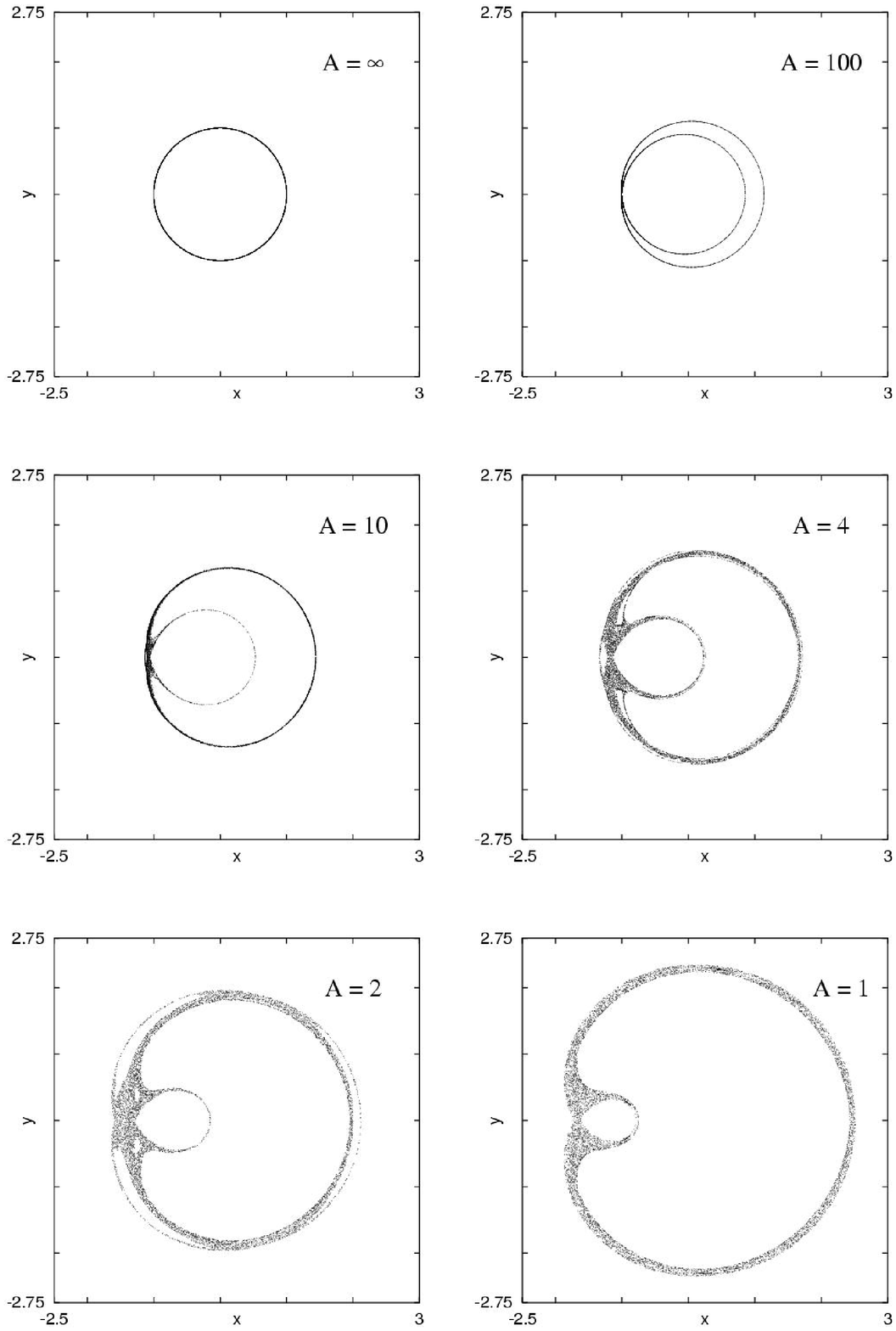}
\caption{
Stroboscopic maps of the period $2 \pi$ for the model described by Eqs.\ (\ref{m3}) and (\ref{m10}). In each case the time interval is $0 < t < 50000$ with the time step of $\Delta t = 0.001$.
}
\label{f1}
\end{figure}

\begin{figure}
\centering
\includegraphics[scale=0.75]{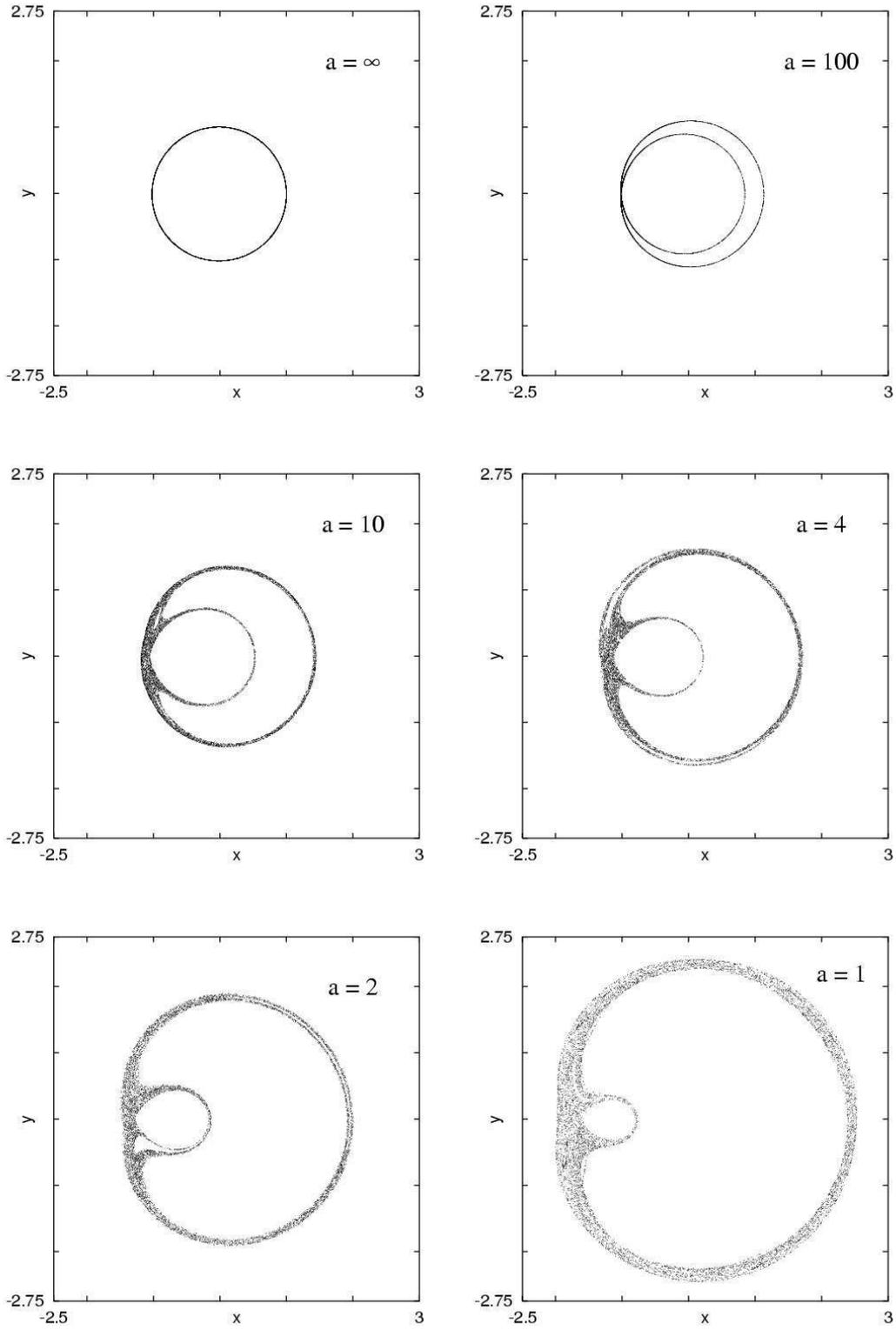}
\caption{
Stroboscopic maps of the period $2 \pi$ for the isospectral model described by Eqs.\ (\ref{m8}) and (\ref{m11}) with $a_0 = a_1 = a$ and $\lambda = \mu = 20$. The time interval and the time step are respectively $0 < t < 50000$ and $\Delta t = 0.005$.
}
\label{f3}
\end{figure}

In this article we have discussed the set of equations that could serve as the reference one for chaotic causal trajectories and revealing the way in which the causal orbits become chaotic ones. This work is the first attempt to contribute to this difficult problem. The results presented here seem to suggest that the way from a regular to chaotic behaviour leads via the formation and break-up of a homoclinic orbit. Unfortunately, the formal derivation of the orbit and of the critical points is possible only for the model of Section 3 above and for the details we refer the reader to \citep{b15}. In other cases we are able to do that only numerically. To emphasize this, we have prepared sequences of pictures in Figs.\ \ref{f3} and \ref{f4} clearly showing their close similarity to those in Fig.\ \ref{f1}. It is indeed common for the way the chaotic behaviour appears in Fig.\ \ref{f1} to be essentially independent of the model that is used to generate it. Results of our paper thus bridge the gap between the theory of dynamical systems and the causal quantum trajectories.

We can consider the sequence of pictures in Fig.\ \ref{f1}  as a possible way leading to the appearance of chaotic orbits in the Bohmian, or equivalently, in the hydrodynamical formulation of quantum mechanics. Quite strong support of the conjecture is in our opinion convincingly manifested in the series of plots in the above figures. What we have discussed here has not been observed so far since the models under consideration were too complex and the properties of trajectories depicted here were masked in their very complicated dynamics.

The proof that some orbits in Fig.\ \ref{f1}, and hence in Figs.\ \ref{f3} and \ref{f4} as well, are chaotic ones, follows directly from the results of our recent paper \citep{b15}. That is why we have calculated the largest Lyapunow exponent just for one orbit with $A = 2$ to show what the typical order of magnitude for the quantity and models under consideration is.

Looking for the simplest yet non-trivial dynamics systems, both conservative and dissipative ones, has attracted much interest for a number of years. Representative studies on the subject together with the lists of such models can be found in \citep{b20,b21}. Our work is a contribution in this field for a special class of dynamical systems generated by the guidance equation (\ref{m1}) of the de Broglie-Bohm quantum mechanics. The model based on Eqs.\ (\ref{m3}) and (\ref{m10}) can be considered as the simplest non-trivial one for the reasons outlined throughout the paper. The name is additionally justified by the fact that the wave function we have used has only one moving node. Our numerical experiments with a number of other models lead to the conjecture that for the systems of two non-autonomous equations of the form of Eq.\ (\ref{m1}), existence of at least one moving node seems to be a necessary condition for chaotic orbits to exist. In our model the node moves along a circle in the integrable case and along an ellipse as the system is chaotic.

We should emphasize at this point that a huge number of dynamical systems can be generated with help of Eq.\ (\ref{m1}) since they are a functional of the used wave function. The resulting systems of two non-autonomous equations, necessary by the well-known Poincar\'e-Bendixson theorem for chaotic behaviour to appear, are not similar to any of systems studied so far. Our results show, however, that for the particular choice of the velocity field generator, \ie the wave function, we can observe mechanism of obtaining chaotic orbits similar to the one known for the systems forced by the disturbance periodic in time. To find such a behaviour for ``the quantum trajectories'' of Eq.\ (\ref{m1}) we were constrained to a specific choice of functions without a particular physical significance. The similarity of Bohmian systems to the well-known dynamical ones is presumably still preserved for physically important wave functions. This statement is, for now, an open question.

\section*{Acknowledgments}

This work has been supported in part by The Polish Government (KBN Grant No.\ 2 P03B 121 16).

\end{document}